\documentstyle[prl,aps,epsf]{revtex}
\begin{document}
\draft
\twocolumn[\hsize\textwidth\columnwidth\hsize\csname@twocolumnfalse\endcsname

\title{Evidence for weak itinerant long-range magnetic correlations 
in UGe$_2$}
\author{A. Yaouanc,$^1$ P. Dalmas de R\'eotier,$^1$ P.C.M. Gubbens,$^2$ 
C.T.~Kaiser,$^2$ A.A. Menovsky$^3$, M. Mihalik,$^3$ and S.P.~Cottrell$^4$}

\address{$^1$Commissariat \`a l'Energie Atomique, D\'epartement de Recherche
Fondamentale sur la Mati\`ere Condens\'ee \\
F-38054 Grenoble Cedex 9, France\\
$^2$Interfacultair Reactor Instituut, Delft University of Technology,
2629 JB Delft, The Netherlands\\
$^3$Van der Waals-Zeeman Instituut, Universiteit van Amsterdam, 
1018 XE Amsterdam, The Netherlands\\
$^4$ISIS Facility, Rutherford Appleton Laboratory, Chilton, Didcot,
OX11 0QX, UK}

\date{\today} \maketitle 

\begin{abstract}
Positive muon spin relaxation measurements performed on the ferromagnet 
UGe$_2$ reveal, in addition to the well known localized $5f$-electron density 
responsible for the bulk magnetic properties, the existence of itinerant 
quasi-static magnetic correlations. Their critical dynamics is well 
described by the conventional dipolar Heisenberg model. These correlations 
involve small magnetic moments. 
\end{abstract}
]
The discovery of superconductivity below 1 K within a limited pressure range 
in the ferromagnet UGe$_2$ \cite{Saxena00,Tateiwa01,Huxley01,Bauer01} provides 
an unanticipated example of coexistence of superconductivity and strong 
ferromagnetism. The electronic pairing mechanism needed for superconductivity 
is believed to be magnetic in origin. However, it is amazing that  
ferromagnetically ordered uranium magnetic moments with so large magnitude
($\sim 1.4 \, \mu_{\rm B}$ at ambient pressure as deduced from 
magnetization measurements) are directly involved \cite{Ginsburg57}. 
Since the pairing must involve the conduction electrons, it is important to
characterize their magnetic properties. Because of the restrictions imposed
by the magnetic form factor, this can not be done by diffraction 
techniques.
As the muons localize in interstitial sites, they have the potentiality 
to yield information on the conduction electrons.
Here we show, using the muon spin relaxation technique, that UGe$_2$ 
is actually a dual system where two sub-states of $f$ electrons coexist. 
We indeed report the existence at ambient pressure of itinerant long-range 
magnetic correlations with magnetic moments of 
$\sim 0.02 \, \mu_{\rm B}$ and a spectral weight in the megahertz 
range. A quantitative understanding of this state is moreover reached 
assuming that these correlations involve only long wavelength fluctuation 
modes.
 
UGe$_2$ is a ferromagnet with a Curie temperature $T_{\rm C} \simeq 52$ K 
which crystallizes in the orthorhombic ZrGa$_2$ crystal 
structure (space group Cmmm) \cite{Oikawa96,Boulet97}. Magnetic 
measurements indicate a strong magnetocrystalline anisotropy 
\cite{Menovsky83,Onuki92,Huxley01} with easy magnetization axis along the 
${\bf a}$ axis. 

We present results obtained by the muon spin relaxation ($\mu$SR) 
technique. Fully polarized muons are implanted
into the studied sample. Their spin (1/2) evolves in the local magnetic field, 
${\bf B}_{\rm loc}$, until they decay into positrons. Since the positron
is emitted preferentially in the direction of the muon spin at the decay time,
it is possible to follow the evolution of the muon spin polarization
\cite{Dalmas97,Amato97}. The measured physical parameter  
is the so-called asymmetry which characterizes the anisotropy of the positron
emission. Below $T_{\rm C}$, if ${\bf B}_{\rm loc}$ has a component 
perpendicular to the initial 
muon beam polarization, ${\bf S}_\mu$ (taken parallel to $Z$), we expect the 
asymmetry to display spontaneous oscillations with an amplitude maximum for 
${\bf B}_{\rm loc} \perp {\bf S}_\mu$. On the other hand, if 
${\bf B}_{\rm loc} \parallel {\bf S}_\mu$, the asymmetry can be written as
the product of an initial asymmetry related to 
sample, $a_{\rm s}$, and the muon spin relaxation function, $P_Z(t)$, which 
monitors the dynamics of ${\bf B}_{\rm loc}$. 

UGe$_2$ crystals were grown from a polycrystalline ingot using a Czochralski
tri-arc technique \cite{Menovsky83}.
We present results for two samples. Each consists of pieces cut from the 
crystals, put together to form a disk and glued on a 
silver backing plate. They differ by the orientation (either parallel or 
perpendicular) of the ${\bf a}$ axis relative to the normal to the 
sample plane.
The measurements were performed at the EMU spectrometer of the ISIS facility,
from 5 K up to 200 K, mostly in zero-field. 
Additional $\mu$SR spectra were recorded with a longitudinal field.

We found that the temperature dependence of $a_{\rm s}$ for 
${\bf S}_\mu \parallel {\bf a}$ is consistent with
${\bf B}_{\rm loc} \parallel {\bf a}$. In agreement with that conclusion, 
a spontaneous muon spin precession resulting in wiggles in the asymmetry
is observed for ${\bf S}_\mu \perp {\bf a}$. Defining 
$T_{\rm C}$ as the temperature at which the
wiggles disappear, we found $T_{\rm C}$ = 52.49 (2) K. This value coincides
with the maximum of the relaxation rate (to be evidenced below) for 
${\bf S}_\mu \parallel {\bf a} $ and ${\bf S}_\mu  \perp {\bf a} $. 

In this letter we 
focus on the description of data taken around the Curie point.

\begin{figure}
\begin{center}
\epsfxsize=82 mm
\epsfbox{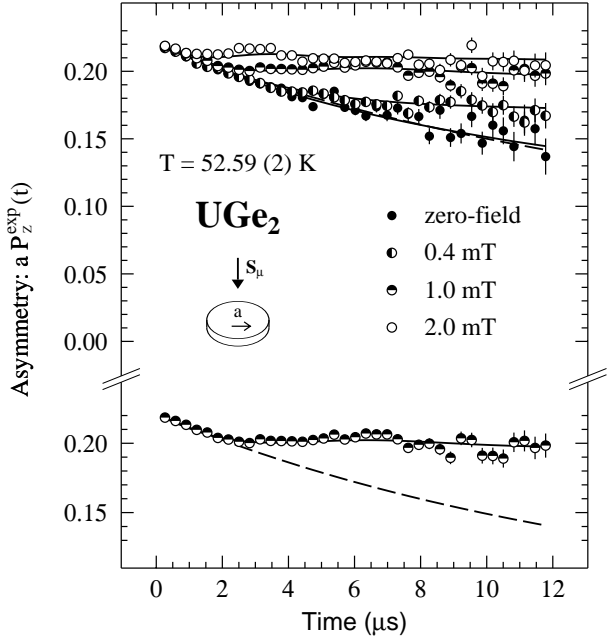}
\end{center}
\caption{Upper panel: examples of $\mu$SR spectra recorded in zero and
longitudinal 
fields at $T$ = 52.59 (2) K (above $T_{\rm C} = 52.49$ (2) K) for 
${\bf S}_\mu \perp {\bf a}$. The solid lines are fits assuming a 
squared-Lorentzian distribution for the modulus of the field at the muon 
site, $B_{\rm loc}$. The dashed line, 
which is the result of the fit of the zero-field spectrum with an exponential 
relaxation function, can not be distinguished from the solid line except above 
$\sim$ 11 $\mu$s. In the lower panel, the comparison of the 1.0 mT spectrum 
with the prediction of an exponential fit shows that this model is not
valid in longitudinal fields. The field dependence at small field of 
$P_Z(t)$ proves that the field distribution at the muon site is 
quasi-static.}    
\label{fig_spectra}
\end{figure}

All the spectra were analyzed as 
a sum of two components: $a P_Z^{\rm exp}(t) = a_{\rm s} P_Z(t) + a_{\rm bg}$.
The first component describes the $\mu$SR signal from the sample and the
second accounts for the muons stopped in the background, {\sl i.e.} the 
cryostat walls and sample holder. In zero-field, for all relevant 
temperatures and for the two orientations of ${\bf S}_\mu$ relative to 
${\bf a}$, $P_Z(t)$ is well described by an exponential function: 
$P_Z(t)$ = $\exp(- \lambda_Z t)$ where $\lambda_Z$ measures 
the spin-lattice relaxation rate at the muon site. An example is shown in 
Fig.~\ref{fig_spectra}. 
$a_{\rm bg}$, which is basically temperature
independent, was measured for ${\bf S}_\mu \perp {\bf a}$ and 
$T< T_{\rm C}$ as the constant background signal. We got $a_{\rm bg}$ = 0.077 
\cite{notebg}. For ${\bf S}_\mu \parallel {\bf a}$, it could only be 
estimated from the sample size 
since the relaxation was never strong enough to measure it directly. We took
$a_{\rm bg}$ = 0.064. The uncertainty on this $a_{\rm bg}$ leads to an 
uncertainty on the absolute value of  
$\lambda_Z({\bf S}_\mu \parallel {\bf a})$ of $\sim 10 \%$. 

\begin{figure}
\begin{center}
\epsfxsize=82 mm
\epsfbox{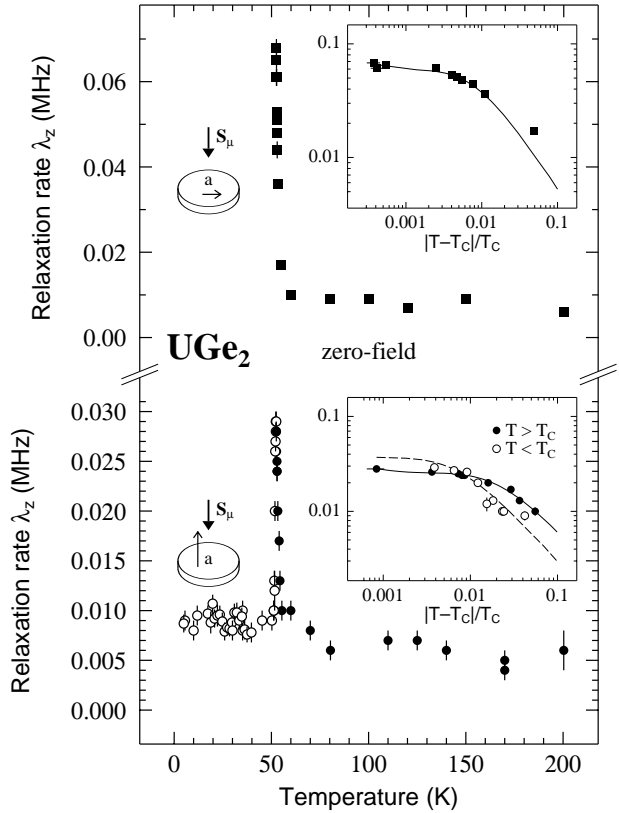}
\end{center}
\caption{Temperature dependence of $\lambda_Z$ measured in zero-field for
${\bf S}_\mu \perp {\bf a}$ and ${\bf S}_\mu \parallel {\bf a}$ in the upper
and lower panels, respectively. The inserts display $\lambda_Z(T)$ near
$T_{\rm C}$. The solid and dashed lines are the results of fits 
for a dipolar Heisenberg ferromagnet as explained in the main text.
Since ${\bf B}_{\rm loc} \parallel {\bf a}$, we can not observe a 
spin-lattice relaxation process for ${\bf S}_\mu \perp {\bf a}$ in the 
ordered state. The point for ${\bf S}_\mu \perp {\bf a}$ at 
$(T - T_{\rm C})/T_{\rm C} = 0.05$ does not fit the critical 
description, pointing out that it was recorded outside the critical region
for that geometrical configuration. The behavior of $\lambda_Z(T)$ near 
$T_{\rm C}$ is typical for a ferromagnet in its critical regime.}    
\label{lambda}
\end{figure}

In Fig.~\ref{lambda} we display $\lambda_Z(T)$ measured in zero-field for 
${\bf S}_\mu \perp {\bf a}$ and ${\bf S}_\mu \parallel {\bf a}$. For both 
geometries, $\lambda_Z(T)$ exhibits a maximum at $T_{\rm C}$. It is due to the
critical slowing down of the spin dynamics. Surprisingly the anisotropy 
between the orientations is very weak although UGe$_2$ is known to be 
extremely anisotropic \cite{noteani}. Furthermore we show 
in the following lines that $\lambda_Z(T)$ near $T_C$ is quantitatively 
understood in the framework of the Heisenberg model with dipolar 
interactions, whereas UGe$_2$ is considered as an Ising system. The 
magnetic signal that we observe has therefore a different
origin than the well documented uranium magnetic state observed e.g. by 
macroscopic measurements. 

$\lambda_Z(T)$ has been computed several years ago 
\cite{Yaouanc93} for the critical regime 
of dipolar Heisenberg ferromagnets and has been successfully compared to
experiments \cite{Yaouanc93,Yaouanc96,Henneberger99}. It is based on the 
derivation of the static and dynamical scaling laws from mode coupling
theory \cite{Frey94}. The two scaling variables
at play depend on two material parameters: $\xi_0$, the magnetic correlation
length at $T=2 T_{\rm C}$, and $q_D$, the dipolar wave vector which is a 
measure of the strength of the exchange interaction relative to the dipolar
energy. This model initially derived for the paramagnetic phase applies
also below $T_C$ \cite{Yaouanc96}.

Specifically, the model predicts that $\lambda_Z(T)$ = 
${\cal W}[a_{\rm L} I^{\rm L}(T) + a_{\rm T} I^{\rm T}(T)]$ where 
$I^{\rm {L,T}}$ \cite{Dalmas94} are scaling functions obtained from mode 
coupling theory and $a_{\rm {L,T}}$  are parameters determining respectively 
the amount of longitudinal (L) and transverse (T)
fluctuations probed by the measurements. The L,T indices denote the 
orientation relative to the wave vector of the fluctuation mode.
$a_{\rm L,T}$ only depend on muon site properties.
The result of the fit of $\lambda_Z(T)$ is shown in the inserts of 
Fig.~\ref{lambda}. The divergence of $\lambda_Z$ at $T_{\rm C}$ is strongly 
reduced by the effect of the dipolar interaction \cite{Frey94}.
The temperature scale gives the product $q_{\rm D} \xi_0$
\cite{Yaouanc93}.
For ${\bf S}_\mu \perp {\bf a}$, we get $q_{\rm D} \xi_0$ = 0.021 (2), and for
${\bf S}_\mu \parallel {\bf a}$, $q_{\rm D} \xi_0^+$ = 0.043 (2) and 
$q_{\rm D} \xi_0^-$ = 0.020 (2). The index $+$ ($-$) on $\xi_0$ specifies 
that we consider the paramagnetic (ferromagnetic) state. 
$ \xi_0({\bf S}_\mu \parallel {\bf a}) >
 \xi_0({\bf S}_\mu \perp {\bf a})$ in the paramagnetic state, suggesting that
the magnetic correlations are somewhat anisotropic. The fact that 
$\xi_0^+ > \xi_0^-$ is an expected feature \cite{Dalmas97}.
The relaxation rate scale yields  ${\cal W}^+ a_{\rm L}$ = 
0.140 (4) MHz and ${\cal W}^- a_{\rm L}$ = 0.20 (2) MHz for 
${\bf S}_\mu \parallel {\bf a}$.
The transverse contribution to 
$\lambda_Z$ for both $T < T_{\rm C}$ and $T > T_{\rm C}$ is more
difficult to estimate since $a_{\rm T}$ is found much lower than $a_{\rm L}$.
Reasonable fits are obtained with $a_{\rm T}/a_{\rm L}$ = 0.036 (14).
We have computed $a_{\rm L}$ and $a_{\rm T}$ for different possible
muon sites and found only one site satisfying $a_{\rm T} < a_{\rm L}/2$. 
This is site $2b$ (in Wyckoff notation) of coordinates (0, 1/2, 0) for which
$a_{\rm L}$ = 1.2486, $a_{\rm T}$ = 0.0386. We then deduce 
${\cal W}^+$ = 0.112 (3) MHz and ${\cal W}^-$ = 0.161 (16) MHz. 
The scale deduced from 
the measurements with ${\bf S}_\mu \perp a$ is about twice as large, 
pointing out again to the weak anisotropy of the magnetic correlations.

In order to further characterize the relaxation near $T_{\rm C}$, we 
performed at a given temperature longitudinal field 
measurements for the two orientations of ${\bf S}_\mu$ relative to ${\bf a}$. 
The field responses for the two geometries are similar.
An illustration is given in Fig.~\ref{fig_spectra}. Surprisingly, the spectra
are field dependent at extremely low $B_{\rm ext}$, proving
that the probed magnetic fluctuations are quasi-static 
(fluctuation rate in the MHz range) and since $\lambda_Z$ is small, 
the associated magnetic moment must be small as well. Quantitatively, the 
field dependence of $P_Z(t)$ can not be described consistently either by a 
simple exponential relaxation form (see the lower panel of 
Fig.~\ref{fig_spectra}) 
nor by a relaxation function computed with the strong collision model 
assuming an isotropic Gaussian component field distribution 
\cite{Hayano79}. On the other hand, the relaxation is well explained if we 
assume that the distribution of the local field at the muon, 
$B_{\rm loc}$, is squared-Lorentzian 
\cite{Walker80}. We write $P_Z(t)$ = $P_Z(\Delta_{\rm Lor}, \nu_f ,t)$ where 
$\Delta_{\rm Lor}$ characterizes the width of the field 
distribution and $\nu_f$ its fluctuation rate \cite{Uemura85}. A global 
fit of the spectra ($B_{\rm ext}$ = 0, 0.2, 0.4, 0.6, 0.8, 1.0
and 2.0 mT) taken at a given temperature is possible. For 
${\bf S}_\mu \perp {\bf a}$ at $T$ = 52.59 (2) K, the description of the seven 
spectra is done with $\Delta_{\rm Lor}$ = 70 $\mu$T and $\nu_f$ = 0.10 MHz.
For ${\bf S}_\mu \parallel {\bf a}$ at $T$ = 52.47 (2) K the two parameters 
are $\Delta_{\rm Lor}$ = 40 $\mu$T and $\nu_f$ = 0.50 MHz: the
zero-field spectra have therefore been recorded in the motional narrowing limit
($\nu_f/ (\gamma_\mu \Delta_{\rm Lor}) > 1$ where $\gamma_\mu$ is the 
muon gyromagnetic ratio; $\gamma_\mu$ = 851.6 Mrad s$^{-1}$T$^{-1}$). 
This justifies the formalism used to treat $\lambda_Z(T)$ close to $T_{\rm C}$.

We now present an interpretation of our results.

We first note that the detected fluctuations can not arise directly from the 
localized uranium $5f$ electrons since $\nu_f$ would then be in the THz 
window as estimated from $\nu_f \simeq k_{\rm B} T_{\rm C} /\hbar$, rather
than in the MHz range as measured. We also already mentioned that 
the observed $\mu$SR signal has not the properties expected from the known
macroscopic properties.  These apparently conflicting results can be 
understood if the $5f$ electrons are viewed as two electron subsets.  
This picture has already been argued for UCu$_5$ \cite{Schenck90} and
UPd$_2$Al$_3$ \cite{Caspary93,Feyerherm94,Metoki98,Bernhoeft98,Sato01}. 
However, for UGe$_2$ the signatures of both subsets are found at 
a single temperature, the Curie temperature, whereas for UCu$_5$ and 
UPd$_2$Al$_3$ the temperatures at which the two subsets
are detected are far apart. So UGe$_2$ presents a novel variant of the
two electron subset model.
Within this picture, the anisotropy of the magnetization
arises from the localized $5f$ spectral density and the magnetic fluctuations
probed by $\mu$SR is a signature of the band-like electrons. We do not detect
the signature of the localized $5f$ electrons, because of the strong motional 
narrowing of the related relaxation rate. It is observed for the ferromagnet 
YbNiSn; see Ref. \onlinecite{Yaouanc99}. 

The effect of the dipolar interaction on the quasi-elastic 
linewidth, $\Gamma (q)$, of the fluctuations has already been observed 
for the weak itinerant ferromagnet Ni$_3$Al \cite{Semadeni00}. 
In particular, at criticality $\Gamma (q) \propto q^{5/2}$, as expected from 
scaling \cite{Frey94}. Thus it 
is not completely surprising to detect its influence on $\lambda_Z(T)$ for
the band-like electrons of UGe$_2$. Quantitatively, the data have been 
described in the established framework of critical dynamics 
\onlinecite{Frey94}. We shall now prove that the 
detected magnetization density arises entirely from long wavelength, {\sl i.e.}
small $q$, fluctuations. The magnetic properties of weak 
itinerant ferromagnets are explained  with the latter hypothesis 
\cite{Lonzarich85,Moriya85}. In our model the values of ${\cal W}$ and 
$\nu_f$ and of the magnitude of the band-like uranium magnetic 
moment, $m_{\rm U}$, are controlled by two wavevectors: 
$q_{\rm D}$ already introduced and the cut-off wavevector, $q_c$, 
which sets the upper bound for the wavevector of the fluctuations involved 
in the build-up of the magnetization density. For simplicity we
consider that the magnetic properties of this electronic subset are isotropic.
We shall detail the analysis of the data 
taken with ${\bf S}_\mu \parallel {\bf a}$. The same approach works equally 
well for the data recorded with ${\bf S}_\mu \perp {\bf a}$. 
As explained below, we get an overall consistent picture setting
$q_{\rm D} = 1.0 \, \times 10^{-3}$ \AA$^{-1}$ and $q_c  = 0.1 $ \AA$^{-1}$.

The magnetization arising from the conduction electrons can be viewed as a 
stochastic variable with a variance $\langle (\delta{\cal M})^2 \rangle$. 
From the fluctuation-dissipation (Nyquist's) theorem \onlinecite{White70}, 
$\langle (\delta{\cal M})^2 \rangle$ obeys the sum rule 
\begin{eqnarray}
\langle (\delta{\cal M})^2 \rangle = {3 \, k_{\rm B} T \over 2\pi^2\mu_0}
\int_0^{q_c} \chi (q) q^2 {\rm d}q,
\label{mag}
\end{eqnarray}
if the energy of the magnetic fluctuations is smaller than the thermal energy.
$\mu_0$ is the permeability of free space. Assuming an Ornstein-Zernike 
form for the wavevector dependent susceptibility, $\chi (q)$, 
and since $q_{\rm D}$ is very small, 
$\langle (\delta{\cal M})^2 \rangle \simeq$ 
$3\, k_{\rm B} T \, q_{\rm D}^2\, q_c / (2 \pi^2 \mu_0)$. Since 
$m_{\rm U} = v_0 \sqrt{\langle (\delta{\cal M})^2 \rangle }$ where $v_0$ is 
the volume per uranium atom ($v_0$ = 61.6 \AA$^3 $), we infer 
$m_{\rm U} = 0.02 \, \mu_{\rm B}$ at $T_{\rm C}$. 
Interestingly, the analysis of polarized neutron scattering data suggests 
for the conduction electrons a magnetic moment of 
$0.04\,(3) \, \mu_{\rm B}$ at low temperature \cite{neutron}.

The scale ${\cal W}$ for $\lambda_Z$ can then be computed within the framework
presented above. Numerically, from
Eq. 5.10c of Ref. \onlinecite{Yaouanc93}, we get   
${\cal W}$ = 0.16 MHz, close to the measured values. With the 
same theory, $\hbar \Gamma (q)  = \Omega q^{5/2}$ with 
$\Omega$ = 18 meV\AA$^{2.5}$
at criticality and for small $q_{\rm D}$ 
(see Eq. 4.14b of Ref. \onlinecite{Yaouanc93}).
Since the measured dynamics is mainly driven by the fluctuations at $q_{\rm D}$
\cite{Dalmas94}, we estimate $\nu_f \simeq \Gamma (q_{\rm D})$ =
0.87 MHz, not far from the measured value. 

We now discuss the magnitude of $\Delta_{\rm Lor}$. If the distribution of 
$B_{\rm loc}$ was Gaussian, 
the zero-field width of the distribution would be 
$\Delta_{\rm Gauss}$ = 1.7 mT for muon at site $2b$, and 
$m_{\rm U} = 0.02 \, \mu_{\rm B}$ computed using the Van Vleck type 
formalism of 
Ref.\cite{Hayano79}. However the distribution is squared-Lorentzian rather 
than Gaussian. Such a distribution is observed in systems with 
diluted and disordered magnetic moments \cite{Walker80}. According to 
Uemura {\sl et al.} \onlinecite{Uemura85}, 
$\Delta_{\rm Lor} = \sqrt{\pi/2} \, c \, \Delta_{\rm Gauss}$ where $c$ is the 
concentration of moments at the origin of the distribution. This relation 
leads to $c$ = 1.9 \%, consistent with the usual fact that a tiny fraction 
of the total number of valence electrons are able to contribute to the 
magnetic susceptibility. 

From the $q_{\rm D}$ value we derive the exchange interaction. We obtain
$2 {\cal J} = k_{\rm B} T_{\rm C} /4.2$
(see Eq. 4.4b of 
Ref. \onlinecite{Yaouanc93} and Ref. \onlinecite{Frey94}). 
For comparison, the same method gives $2 {\cal J}/ k_{\rm B} T_{\rm C} = 1/11$ 
and $1/20$ for metallic Fe and Ni, respectively. Therefore
the evaluation of the exchange energy is quite reasonable.
From the measured product $q_{\rm D}\xi_0^+$,
we get $\xi_0^+ \simeq 43$ \AA. This means that the interaction between
the itinerant magnetic moments is relatively long-range, even far outside the 
critical regime. Although about an order of magnitude larger than for 
conventional ferromagnets, $\xi_0^+ $ compares favorably with the neutron 
result for Ni$_3$Al: $\xi_0^+$ = 24 (9) \AA \cite{Semadeni00}. For the same 
compound, we derive from \onlinecite{Lonzarich85} that 
$q_c$ = 0.2~\AA$^{-1} $, a value twice as large as found for UGe$_2$. The 
moment carried by the itinerant electrons is about four times smaller for 
UGe$_2$ than for Ni$_3$Al \cite{Semwal99}. 
Nearest neighbor U atoms form zigzag chains parallel to ${\bf a}$
\cite{Huxley01}. This may lead to magnetic frustration and 
thus explains the disordered nature of the distribution of $B_{\rm loc}$.
A proper understanding of the origin of the squared-Lorentzian distribution 
requires more work.

One may question the uniqueness of our interpretation.
We first note that the observed $\lambda_Z$ can not arise from an impurity 
phase since the measured critical dynamics occurs right at the 
well known Curie temperature of UGe$_2$. 
It could be argued that the observed signal is
the signature of a weak disorder in the uranium magnetic moments.
This has already been seen in UAs \cite{Asch89} where the $\mu$SR signal
below the N\'eel temperature has been attributed to a diluted source of 
small magnetic moments. Their quasi-static nature is related to the absence 
of spin excitations. However, the moments we observe in UGe$_2$ 
are quasi-static even above $T_{\rm C}$.

In conclusion we have shown that at ambient pressure UGe$_2$ is a dual system 
where an electronic subset of itinerant states coexist with the subset of 
localized $5f$ electrons responsible for the well known bulk magnetic 
properties. Its associated magnetic moment is quite small and characterized 
by a very slow spin dynamics. A quantitative picture for that subset is 
achieved by assuming that only fluctuations at long wavelength are at play.
It would be of interest to follow the small moment itinerant state as a 
function of pressure to determine whether the Cooper's pairs arise from it. 
However, it seems difficult to perform that task with $\mu$SR, unless the 
spin-lattice relaxation rate increases appreciably at high pressure.

\end{document}